\documentclass{appolb}
\usepackage{epsfig}
\begin{document}
\title{Recent Developments of A Multi-Phase Transport Model}
\author{Z.W. LIN
\address{Department of Physics, East Carolina University, Greenville
  NC 27858-4353, USA} }
\maketitle
\begin{abstract}
After the public release of A Multi-Phase Transport (AMPT) model in
2004 and detailed descriptions of its physics in a 2005 paper,  
the model has been constantly updated and developed 
to make it more versatile and to include more physical processes. 
This an overview of recent developments of the AMPT model. 
Ongoing work to fix the violation of charge conservation in the code
as well as possible directions for future work are also discussed.
\end{abstract}
  
\section{Introduction}

A Multi-Phase Transport model was constructed specifically for
the study of relativisitc heavy ion collisions. It intends
to serve as a self-contained phenomenological model, 
since it incorporates essential stages of heavy ion collisions from 
the initial condition to final observables on an event-by-event basis, 
including the parton cascade, hadronization and the hadron cascade. 
The default version of the AMPT model was first constructed to predict
particle yields and momentum spectra in heavy ion collisions at RHIC
\cite{Zhang:1999bd}.  
After large elliptic flow was discovered at RHIC, the string melting
version of the AMPT model was introduced \cite{Lin:2001zk} in order to
take into account the effects on flow from the whole partonic system
in the overlap volume (instead of from only minijet partons in the
default version). 

\begin{figure}[h]
\includegraphics[width=\linewidth]{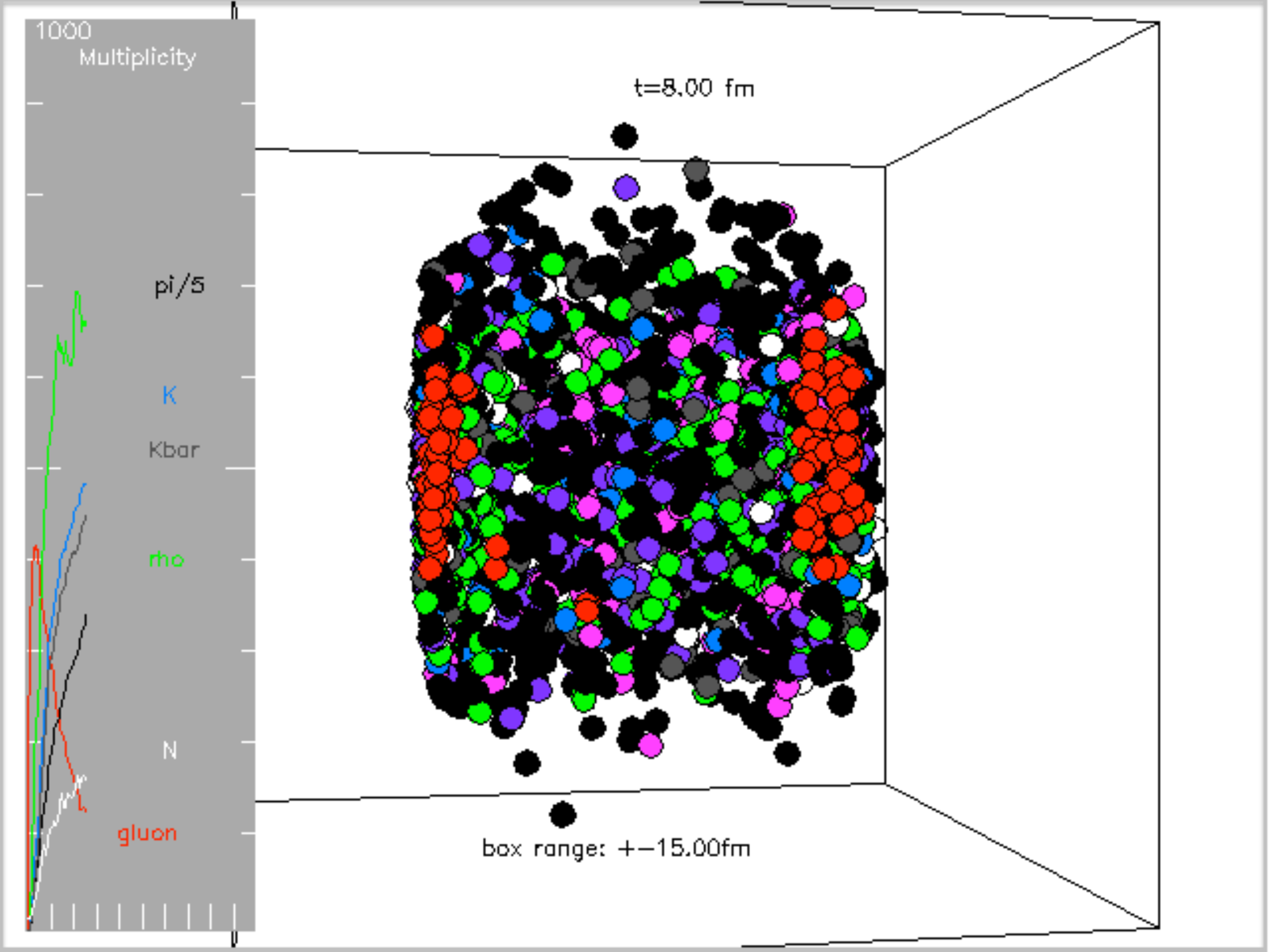}  
\vspace{0.5cm}
\includegraphics[width=\linewidth]{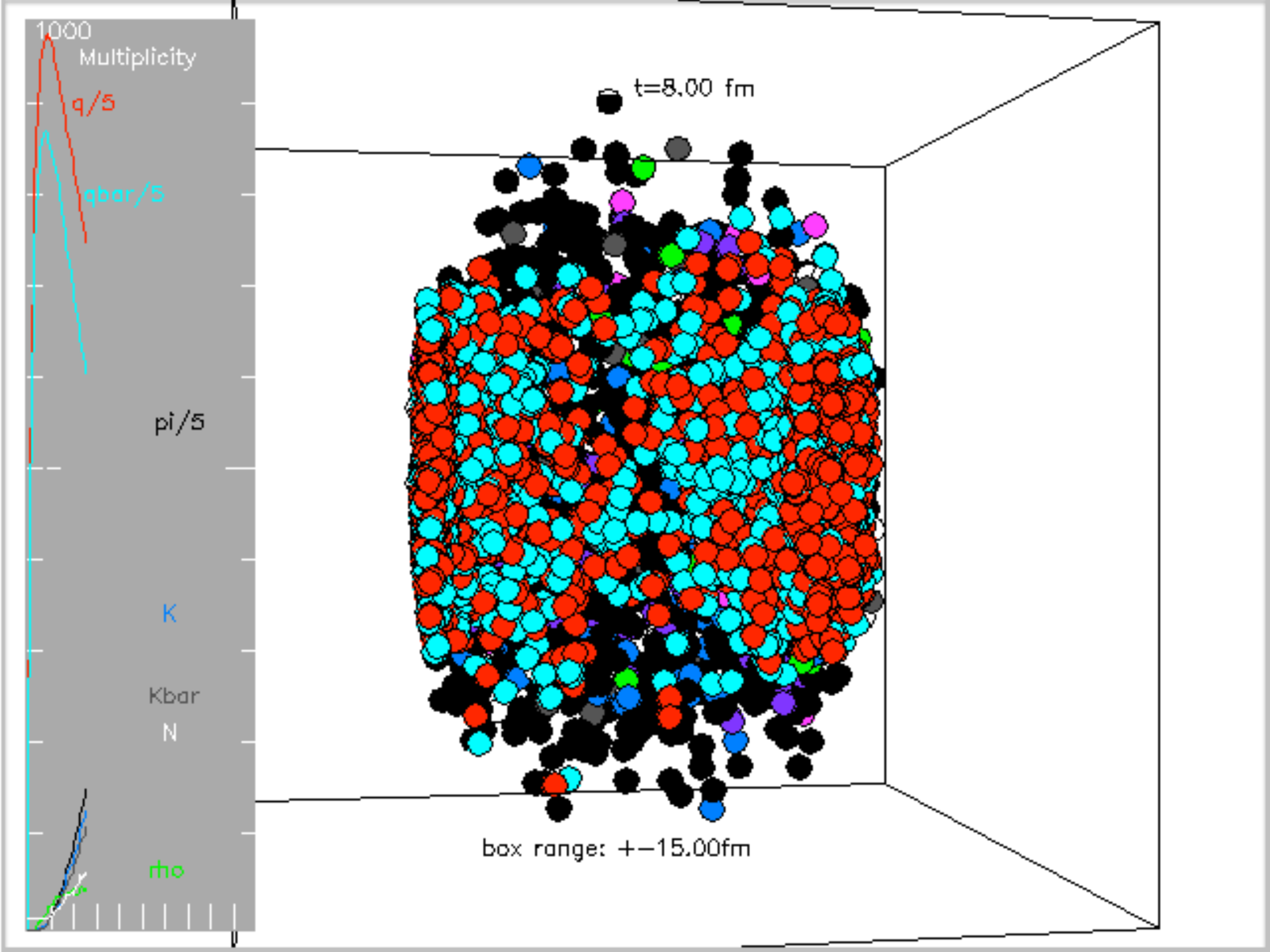}  
\caption{Side view of a 200 AGeV Au+Au event at impact
  parameter $b=0$ fm at the time of 8 fm$/c$.
Upper panel: from the default version of AMPT.
Lower panel: from the string melting version of AMPT (with identical 
initial condition for the event).
}
\label{fig1:def-sm}
\end{figure}

To help illustrate the difference between the default version and the
string melting version of the AMPT model, 
Fig.~\ref{fig1:def-sm} shows the snapshot at the time of 8 fm$/c$ of a
central Au+Au event from the default version (upper panel) and from
the string melting version (lower panel), where both calculations use 
an identical HIJING initial condition such as the position of every
participant nucleon and spectator nucleon. 
This is a side view where the two beams come from the left and right
sides of the showed box respectively, and the full length of the box
is 30 fm along each direction.
The insert on the left side of each panel shows the number of
several particle species in the event as a function of time 
(up to the time of 8 fm$/c$ for this figure).
Black particles here represent pions which number in the left insert
has been scaled down by a factor of 5, red particles represent gluons
(in the upper panel) or quarks (in the lower panel), while cyan
particles in the lower panel represent antiquarks. 
We see in both panels that many partons still exist on the left and 
right sides (at large rapidities) while more hadrons have been formed
in the middle (around mid-rapidity); 
this is a consequence of the time dilation that delays the dynamics at
large rapidities when the global time is used to calculate particle
interactions, as is the case for AMPT. 
In the lower string melting panel, we see many more
partons and a delayed hadron phase compared to the upper panel for the
default version, and this demonstrates the dominance of partons in
early times in the string melting version. The full time evolutions of
these events  are available as animation files at the author's website
\cite{amptLink}.

\begin{figure}[h]
\includegraphics[width=\linewidth]{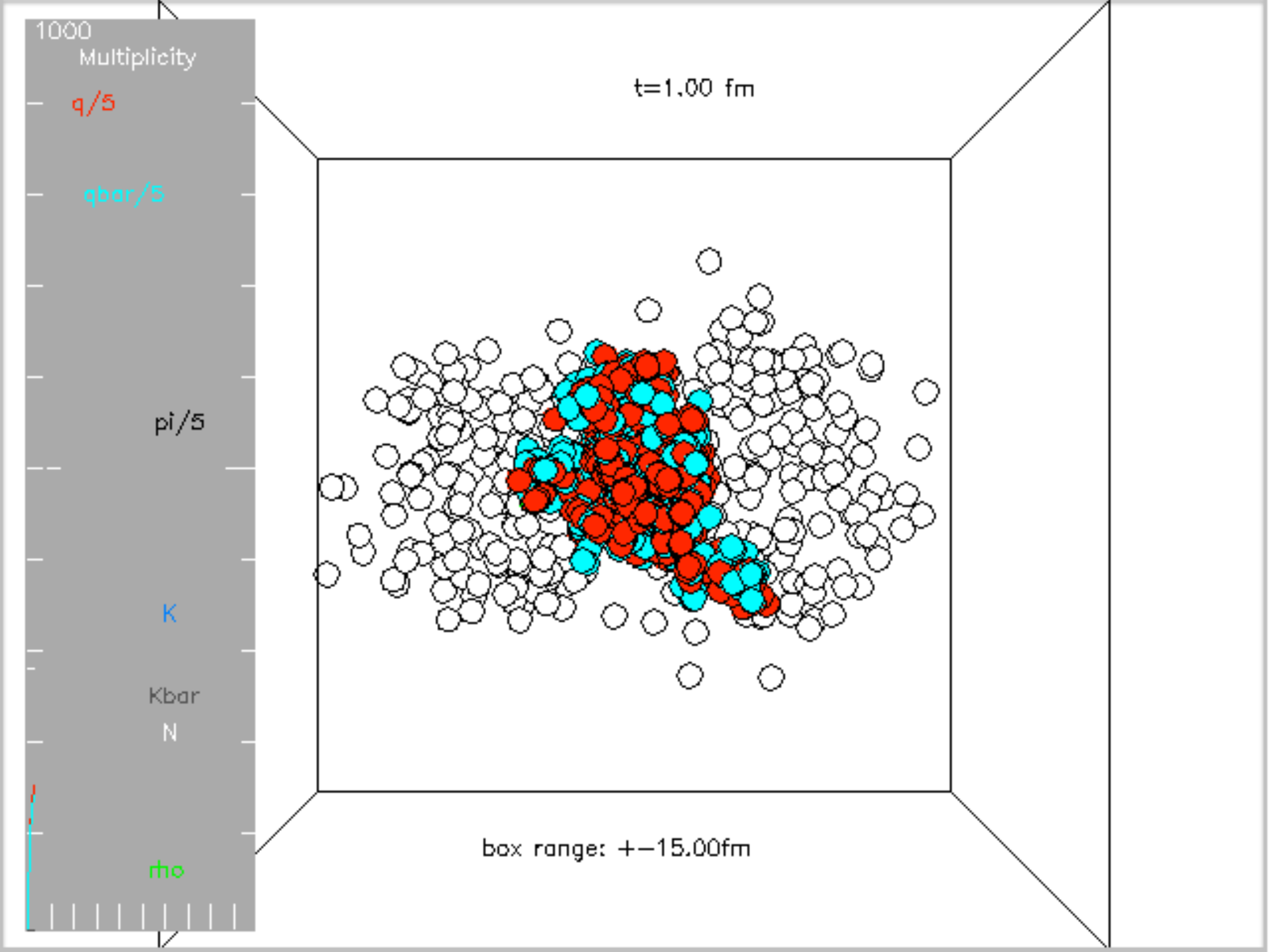}
\caption{View of the transverse plane along the beam axis at 
the time of 1 fm$/c$ for a 200 AGeV Au+Au event at impact parameter
$b=$10 fm from the string melting version of AMPT. 
}
\label{fig2:sm-b10fm}
\end{figure}

The source code of the AMPT model was first publicly released online  
around April 2004, and a subsequent publication \cite{Lin:2004en}  
provided detailed descriptions of the model such as the included physics 
processes and modeling assumptions.  
The model has since been applied to the study of many 
observables in heavy ion collisions. 
The AMPT model is also a useful test bed of different ideas; for
example, the connection between triangular flow and initial
geometrical fluctuations was discovered with AMPT
\cite{Alver:2010gr}. 
Shown in Fig.~\ref{fig2:sm-b10fm} is the view along the beam axis
for the transverse positions of particles in a peripheral Au+Au event  
at the early time of 1 fm$/c$ from the string melting version of AMPT.
Here, white particles that are mostly in the peripheral
region represent spectator nucleons, and colored particles are mostly
partons (quarks and antiquarks) that have already formed. 
We can clearly see that the shape of the parton system in the
overlap region is not elliptical in the transverse plane, and for this
particular event it looks quite like a triangle. 
As found in Ref.\cite{Alver:2010gr}, subsequent partonic and
hadronic scatterings convert the irregular geometry in each event into
particle correlations in momentum space such as triangular flow.

\section{Recent Developments of the AMPT Model}

The ``official'' 2004 version of AMPT v1.11(default)/v2.11(string melting) 
and the 2008 version v1.21(default)/v2.21(string melting) are
available at the OSCAR website \cite{oscar}. 
Since then the AMPT model has been constantly updated and developed,
sometimes at users' requests, 
to make it more versatile and stable and to include more physical processes.
Several of these more recent ``test'' versions of the source codes are 
available at the author's website \cite{amptLink}, where the 
AMPT Users' Guide and the readme file in the source code 
detail the changes made in each version.

The following summarizes some of the developments and updates of the
model after the 2008 version v1.21/v2.21. 
Deuteron(d) interactions have been included in the hadron
cascade, including inelastic collisions $d+M \leftrightarrow B+B$ 
(where $M$ represents a meson and $B$ represents a baryon), elastic
collisions of deuterons, and corresponding interactions for anti-deuterons
\cite{Oh:2009gx}. 
We have also extended the functionality of the code to 
allow users to 
1) insert arbitrary user-defined hadrons at the start
of hadron cascade,
2) obtain the transverse positions of all initial nucleons
(participant and spectator nucleons) 
as well as their present and original flavor codes, 
3) trigger events so that each event will have at least one initial
minijet parton with transverse momentum above a user-defined value, 
4) embed one back-to-back $q/\bar q$ pair per event at the
user-specified transverse momentum and transverse position at the
start of parton cascade,
5) obtain the complete initial parton information before parton
cascade as well as the complete parton collision history, 
6) set the nuclear shadowing strength anywhere between no shadowing
and the default HIJING shadowing.
In addition, we added deformed Uranium-238 
as a type of projectile and/or target nucleus \cite{Haque:2011aa}; 
since this extension was done in parallel with some of the above
developments, it has not yet been incorporated into the posted 
AMPT versions \cite{amptLink}. 

\section{Ongoing Work and Future Directions}

An ongoing work is to implement electric charge conservation in the 
AMPT model so that it can be used to reliably study charge-dependent
variables such as charge fluctuations in heavy ion collisions.
Currently the AMPT model violates charge conservation due to two
reasons. 
First, the hadron cascade has $K^+/K^-$ as explicit particles but not 
$K^0/{\bar K^0}$, so the code changes $K^0$ to $K^+$ and changes 
$\bar K^0$ to $K^-$ before hadron cascade in order to include neutral
kaons in the hadron cascade, and after hadron cascade the code changes
half of the final  $K^+$ into $K^0$ and changes half of $K^-$ to $\bar
K^0$.  This however introduces a violation of charge conservation. 
Secondly, not all hadron reactions or resonance decays in the hadron
cascade of AMPT are implemented for each possible individual 
isospin configuration; 
for some channels the isospin-averaged cross section is used instead 
and the electric charge of each final-state particle is 
selected randomly from all possible values
(without considering the initial-state charge configuration). 
For example, each final-state pion could have $+, 0$, or $-$ charge 
in the reaction $\pi \eta \rightarrow \pi \pi$ in AMPT, 
where for example $\pi^+ \eta \rightarrow \pi^+ \pi^0$ is allowed but
$\pi^+ \eta \rightarrow \pi^+ \pi^-$ should have been forbidden. 
To fix the charge violation problem in the AMPT model, we thus need to 
add $K^0$ and $\bar K^0$ as explicit particles in the hadron cascade; 
we also need to consider each isospin configuration for the 
hadronic channels and use isospin-dependent cross sections.

Regarding future directions of the AMPT model, 
from the author's point of view we may take two different approaches
to further develop such a self-contained phenomenological model for
relativisitc heavy ion collisions.
One approach is to develop key ingredients of the AMPT model
within the same scheme of a partonic phase followed by a hadronic
phase, while the other approach is to couple the AMPT model with 
a hydrodynamic model.

The first approach will include the following developments:
update of parton distribution functions in nuclei including the
nuclear shadowing effect, 
development of the dynamical parton recombination process (e.g. by
using local parton density as the recombination criterion), 
incorporation of inelastic parton reactions, 
and the consideration of gluons in parton recombination.
For example, a study of the AMPT model \cite{Zhang:2008zzk} has shown
that the quark coalescence in the string melting version of AMPT 
starts too late, i.e., it starts at energy densities that are too low
($\ll 1$ GeV/fm$^3$) compared to the equation of state from lattice
QCD. Implementing parton recombination according to local energy
density would make the effective equation of state in the AMPT model
more realistic and also lead to a better success rate of parton
recombination due to the higher parton density at hadronization.

The second approach aims to couple AMPT with a hydrodynamic model, 
and the resulted hybrid model will provide us a more direct link to 
QCD variables and properties. 
The AMPT model will provide the initial condition, including 
the event-by-event fluctuations and correlations, to the
hydrodynamic model. After the hydrodynamic model calculates the
evolution of the dense matter through the QCD phase transition, 
hadron interactions will be taken into account by the hadron cascade of
the AMPT model. This hybrid model will be a self-contained model 
in that it will have self-consistent initial conditions for different
collision systems at different energies,  
and a 3+1D hydrodynamic code will be ideal so that the 
hybrid model can address non-equilibrium evolution 
and observables at different rapidities.

The author acknowledges support from the Helmholtz International
Center for FAIR, where part of the ongoing work was done at the
Institut fuer Theoretische Physik at Goethe University.

\end{document}